\documentclass[
    ,final            
  ]
  {aipproc}

\layoutstyle{8x11double}

\usepackage{booktabs}
\usepackage{multirow}
\usepackage{array}
\newcolumntype{L}[1]{>{\raggedright\let\newline\\\arraybackslash\hspace{0pt}}m{#1}}
\newcolumntype{C}[1]{>{\centering\let\newline\\\arraybackslash\hspace{0pt}}m{#1}}
\newcolumntype{R}[1]{>{\raggedleft\let\newline\\\arraybackslash\hspace{0pt}}m{#1}}

\begin{document}

\title{Rubric Design for Separating the Roles of Open-Ended Assessments}


\author{Leanne Doughty}{
  address={Department of Physics and Astronomy, Michigan State University, East Lansing, MI 48824}
}

\author{Marcos D. Caballero}{
  address={Department of Physics and Astronomy, Michigan State University, East Lansing, MI 48824},
  altaddress={CREATE for STEM Institute, Michigan State University, East Lansing, MI 48824}
}

\begin{abstract}
End-of-course assessments play important roles in the ongoing attempt to improve instruction in physics courses. Comparison of students' performance on assessments before and after instruction gives a measure of student learning. In addition, analysis of students' answers to assessment items provides insight into students' difficulties with specific concepts and practices. While open-ended assessments scored with detailed rubrics provide useful information about student reasoning to researchers, end users need to score students' responses so that they may obtain meaningful feedback on their instruction. One solution that satisfies end users and researchers is a grading rubric that separates scoring student work and uncovering student difficulties. We have constructed a separable rubric for the Colorado Classical Mechanics/Math Methods Instrument that has been used by untrained graders to score the assessment reliably, and by researchers to unpack common student difficulties. Here we present rubric development, measures of inter-rater reliability, and some uncovered student difficulties. 
\end{abstract}

\maketitle


\section{Introduction}
In recent years the PER community has placed an increased emphasis on improving student learning in upper-division physics courses (e.g., \cite{Manogue01,Singh08}).  Assessments are a useful tool for driving and validating course transformations undertaken to achieve this goal. They can be used to identify persistent student difficulties to inform instruction. Furthermore, they provide a measure of student performance, which helps to evaluate the effectiveness of different pedagogies \cite{Pollock12a}.

In introductory physics, standardized assessments like the FCI \cite{Hestenes92}, consisting largely of conceptual multiple-choice questions, have allowed for reliable measurement of student performance across universities and over time (e.g., \cite{Hake98}). They have also been used for identifying specific difficulties students have employing concepts (e.g., \cite{Redish99}). Upper-division courses, however, utilize more sophisticated mathematics and require students to use a broader set of skills and practices to solve problems. While open-ended questions are better suited for assessing students' ability to approach and to solve problems in upper-division, using them to measure student performance reliably so that meaningful comparisons can be made requires significant training for graders \cite{Chasteen12}. 

One solution that would allow open-ended assessments to fulfill both roles is the use of separate rubrics: a grading rubric that untrained graders can use to score students' answers and measure performance; and a difficulties rubric for trained researchers to unpack common student difficulties. We have developed such rubrics for the Colorado Classical Mechanics/Math Methods Instrument (CCMI) \cite{Caballero13}. 

This paper presents the design and a description of the grading and difficulties rubrics. Outcomes from these rubrics will be the focus of a longer publication. To describe the development of the rubrics, we chose to focus on just one question from the CCMI that asks students to construct a differential equation. We will discuss whether the grading rubric for this question gives reliable scores when used by untrained graders, and also what information can be obtained from the difficulties rubric.

\section{Data}

\begin{figure*} [t!]
\small
\fbox{
\begin{minipage}{6.5in}
{\bf{Learning Goal Evaluated: Students should be able to use Newton's laws to translate a given physical situation into a differential equation}}\\

A particle (mass, {\it{m}}) is confined to move on the {\it{x}}-axis between two objects that attract it. The particle does not leave the region between the two attractive objects. 
\begin{itemize}
\item One object is located at  {\it{x}} = 0, and the attractive force between the object and the particle is proportional to the square of the distance between them with proportionality constant {\it{c}}.
\item The second object is located at {\it{x}} = 10, and the attractive force between the object and the particle is inversely proportional to the distance between them with proportionality constant {\it{k}}.
\end{itemize}
Write down a differential equation that describes the position of the particle as a function of time, {\it{x(t)}}.
\end{minipage}}
\caption{CCMI question designed to assess how well students can construct the equations of motion from a description of a physical situation. Responses are open-ended with students providing as much or as little information as they see fit.}
\label{fig:DifferentialQ}
\end{figure*}

Course transformation of a Classical Mechanics/Math Methods course at The University of Colorado (CU) \cite{Pollock12} was initiated with the development of broad course-scale learning goals outlining what instructors wanted students to be able to do at the end of the course (e.g. {\it{translate a physical description to a mathematical equation}}), and topic-scale learning goals that blended concepts and skills (e.g. {\it{students should be able to use Newton's laws to translate a given physical situation into a differential equation}}) \cite{LearningGoals}. The CCMI is an open-ended instrument designed to assess a subset of these topic-scale learning goals \cite{Caballero13}. As such, the instrument highlights specific areas where students struggle. Moreover,  as most course-scale goals are incorporated in at least one question, it also provides an indication of how well the course is meeting its overall goals.

Figure \ref{fig:DifferentialQ} shows the question designed to assess the learning goals mentioned above. Conceptually the question examines students' understanding of the relationship between force and position while also testing how well students are able to translate the physical description into a mathematical expression. 

The data presented in this paper was collected {\it{in situ}} and was part of the validation process of the CCMI. The CCMI was administered at three medium to large enrollment universities (courses with between 25 and 75 students) with physics majors. Students were given the assessment as a paper-based 50 minute in-class test. Students were not told about it in advance and their performance did not count towards their final grade, though instructors explained that they valued the assessment. Think-aloud interviews were conducted with six students at CU. Students worked through all the questions on the CCMI using Livescribe pens so that their written work could be synced with their speech.

\section{Rubric Design}
The development of both the grading and difficulties rubrics is grounded in student work. Analysis of the first data sets focused on examining what students did. Students' solutions and the combination of errors they made were used to infer where their approach broke down and where they experienced specific difficulties. While we cannot know how students are reasoning from looking at their written work, coordinated interviews and observations have given us a stronger idea. 

Patterns in students' answers for each question were identified with the intent of creating categories for a grading rubric similar to that of the Colorado Upper-division Electrostatics (CUE) Diagnostic \cite{Chasteen12}. The CUE grading rubric explicitly defines what points should be assigned to each question for a variety of student responses, while also categorizing student difficulties. For graders to use this rubric consistently, significant training is required. 

Initial feedback from instructors at CU and elsewhere was that they wanted to be able to use the CCMI to quickly score students' answers to compare from semester to semester and against similar implementations at different institutions. As a result, we decided that the grading rubric should use a mastery approach (described below), and that a separate difficulties rubric be designed to provide an organization of students' approaches and errors to enable a faster yet meaningful interpretation of students' answers for trained users. 

\subsection{Grading Rubric}
The grading rubric is structured based on a mastery approach, where only the final answer is considered and points are taken away for errors in that answer. This means that graders need only attend to one part of the students' answers and can score based on obvious features. 

In order to determine the points that should be allocated to each question on the CCMI a group of faculty at CU were asked to rank the questions based on their  perceived importance of the learning goal(s) that the questions assessed. Similarly, once the categories of errors within each question were determined, faculty ranked the severity of those errors so that point deductions could be assigned. It is not important for all graders to agree with our point scheme but rather that they achieve consistent scores using the rubric. 

The grading rubric for the differential equation question (Figure \ref{fig:DifferentialQ}) is shown in Table \ref{tab:GradingRubric}. A correct answer is worth four points and the rubric describes how points are deducted for different errors, providing examples where necessary (it does not list all the possibilities). The illustrative errors are those commonly seen in students' answers. Error types are weighted based on the significance of the error as determined by faculty at CU. It is possible that an answer may have the correct structure but due to the occurrence of multiple mistakes receive a score of zero. However, this has not been the case for the data that we have scored thus far (N=123). Students who approach the problem correctly, i.e. construct force expressions from the description given, but do not write a differential equation receive no credit for their answer. 

To check inter-rater reliability for the grading rubric, a random sample of 25 student answers were scored by three independent untrained graders and their scores were compared. The graders scores agreed for 24 of the 25 student answers. Cohen's kappa was calculated to be 0.95, which indicates `almost perfect' agreement \cite{Landis77}. This suggests that the rubric successfully produces consistent scores when used by untrained graders. 

\begin{table}
\caption{Grading rubric for the CCMI question shown in Figure \ref{fig:DifferentialQ}. It outlines what points should be taken away for the described errors and provides illustrative examples.}  

\footnotesize

\begin{tabular}{L{1.3cm}L{1.5cm}L{3.8cm}}
\toprule

{\bf{Points}} & {\bf{Error Category}} & {\bf{Description/Examples}} \\

\hline

Full credit (4) & Correct & $m\ddot{x}=-cx^2 + \frac{k}{10-x}$ or an equivalent form\\
\hline
Minus 0.5 point & Neglected mass & Mass does not appear in the differential equation\\
\hline
Minus 1 point each (1 max) & Neglected coefficients & Constants (c,k) do not appear in the differential equation\\
\hline
Minus 1 point each (2 max) & Distance dependence error & Any errors in the distance dependence (e.g. $x$ instead of $x^2$ in the first term, 1/$x$ instead of 1/($10-x$) in the second term)\\
\hline
Minus 1 point each (2 max) & Sign error & Sign error in front of either term (e.g. negative sign instead of positive sign for the second term as written above)\\
\hline
No credit (0) & Incorrect & No credit for any other responses \newline $\bullet$ First order equation in $x$ \newline $\bullet$ Differential equation equal to a constant \newline $\bullet$ Not a differential equation \\
\hline
\end{tabular}

\label{tab:GradingRubric}
\end{table}

\subsection{Difficulties Rubric}

The difficulties rubric seeks to capture information about students' difficulties that is lost in the grading rubric due to its mastery approach. The grading rubric for the differential equation question (Table \ref{tab:GradingRubric}) only accounts for two categories observed in students' answers: those with the correct differential equation; and those with a final expression of the correct structure but with errors. Some examples of other expression types seen in students' answers are listed together in the ``no credit'' section of the grading rubric. The difficulties rubric is inclusive of these other answer types where: the final expression is constructed from force terms using descriptions given but is not a second order differential equation; there is no final expression despite constructing force terms from descriptions given; the final expression is a second order differential equation that does not use the description given. Although students who do not write a second order differential equation receive a zero score for their answers, some had the ability to represent the force descriptions in a mathematical expression(s), but could not translate that to a differential equation correctly. Looking more deeply at all categories of students' responses in this way helps us to pinpoint where students' difficulties lie (in this case moving from force to a differential equation), and also provides us with a broader sense of how students solve problems. 

To help us classify students' approaches and group the large variety of errors seen in students' answers, a task analysis \cite{Catrambone1996} was carried out for the  question. The necessary steps for constructing the differential equation for this question are outlined here:
\begin{itemize}
\item
Visualize the scenario presented in the problem
\item
Determine the distance the particle is from each object in terms of $x$
\item
Write expressions for the magnitude of the forces in terms of $x$
\item
Recognize that the forces due to each object are in opposite directions and determine the sign for each force
\item
Write an expression for the net force on the particle
\item
Recognize that the force can be written as $m\ddot x$, making the expression for the net force a differential equation
\end{itemize}

\begin{table}[t!]
\caption{Difficulties rubric designed to categorize the errors in students' responses to the question shown in Figure \ref{fig:DifferentialQ} and help infer difficulties based on their written work}  

\footnotesize

\begin{tabular}{L{1.7cm}L{3.8cm}L{6cm}L{2.8cm}}
\toprule
{\bf{Category}} & {\bf{Claim}} & {\bf{Evidence}} & {\bf{Example}} \\

\hline

\multirow{3}{1.7cm}{\bf{Distance}} & Does not distinguish between distances & Uses same distance value for each force & $m \ddot x =-cx^2+k/x$ \\

                               & Distinguishes between distances incorrectly & Uses different variables for each distance \underline{or} determines the second distance incorrectly using the position of the second object  & $m \ddot x =-cx^2+k/(x-10)$ \underline{or} $m \ddot x =-cx_1^2+k/x_2$  \\ 

\hline       

{\bf{Force terms}} & Misrepresents the force description in the mathematical expression/s & Error in the way distance is used in either or both force terms \underline{and/or} proportionality constants not present in force terms & $m \ddot x =-1/x+1/(10-x)$ \\
                                
\hline
       
{\bf{Sign}} & Determines or represents the direction of the forces incorrectly & No minus sign in the final expression \underline{or} a minus sign in front of the wrong term in the final expression & $m \ddot x = cx^2-k/(10-x)$ \\  

\hline      

\multirow{2}{1.7cm}{\bf{Structure}} & \multirow{3}{3.8cm}{Difficulty representing the force expression as a differential equation} & Mass does not appear in the final expression & $ \ddot x =-cx^2+k/(10-x)$ \\
                                &  &  One side of the expression is written in terms of $x$ and the other in terms of another variable & $m \ddot x =-cr^2+k/(10-r)$ \\                                  
                                &  & Final expression is a first order differential equation \underline{or} a function    & $dx/dt= cx^2+kx$ \underline{or} $x(t)= cx^2+k/x$  \\
                                &  & No final expression & $F_{1}=-cx^2$, $F_2=k/(10-x)$\\
 
\hline    

\multirow{3}{1.7cm}{\bf{Other approaches}} & Recall a differential equation & Write a generic expression not using the description given in the question & $A\ddot{x}+B\dot+x=0$ \\

 & Use terms given in description to write an expression & Terms $c$, $k$, and $x$ appear in the expression in some form & $x(t)=\dot{x}(k-c)$ \\
 
\bottomrule

\label{tab:DifficultiesRubric}                             
\end{tabular}
\end{table}

Although all these steps need to be taken to construct the differential equation they need not necessarily be completed in this order. For example, students may start writing the net force without writing an expression for each individual force first, but in doing so they must consider each individual force. The think-aloud interviews confirm the variance in the order that students consider these steps. 

Due to the nature of this problem, students' written answers often only contain a simple diagram showing the position of the particle and the objects, two separate force expressions and a final expression (differential equation or otherwise). In some cases, students wrote only the final expression. Therefore, for this particular question we make most of our inferences about student difficulties based on the final expression (for other questions on the CCMI we were able to obtain more information from other parts of students' answers). 

The difficulties rubric for the differential equation question is shown in Table \ref{tab:DifficultiesRubric}. The rubric categorizes student difficulties within certain steps of the task analysis, makes claims within those categories, and provides the evidence from students' written work that we are using to make those claims, along with examples for each. 

For the most part, the final expression (or lack thereof) was indicative of specific student difficulties. One example of this is students who use the same value for distance, in most cases $x$, for both forces. These students failed to distinguish between the distance the particle is from each object. This error was even seen frequently in answers where students had drawn a diagram to indicate the position of the objects and the particle. However, for categories like the sign of the forces, the claim is intentionally broad because an error in the sign of the forces is not enough to infer whether students incorrectly determined the direction of the forces or whether they made a mistake representing that direction.

\section{Conclusions and Future Work}
To separate the roles of open-ended assessment, we have designed a grading rubric that can be used to reliably score student answers, and a difficulties rubric that can be used to break down students' answers and categorize errors such that specific claims can be made about areas where students are struggling. 

Because the difficulties rubric was designed based on student written work, there are cases where we find it difficult to pinpoint specific student difficulties from an error in their answers. For example, it is possible that for the sign category in Table \ref{tab:DifficultiesRubric} answers where there is no minus sign in the final expression students have just not considered the direction of the forces. However, to verify claims like these more written data and targeted interviews are required. Future work will focus on refining and verifying the claims made in the difficulties rubric, and will investigate the outcomes of applying both rubrics to student work. \\

The authors would like to thank the members of PERL@MSU and PER@C for their useful comments and suggestions at various stages of this work. We would especially like to thank Steven Pollock for his valuable feedback on this manuscript.

\bibliographystyle{aipproc}  

\bibliography{References}

\IfFileExists{\jobname.bbl}{}
 {\typeout{}
  \typeout{******************************************}
  \typeout{** Please run "bibtex \jobname" to optain}
  \typeout{** the bibliography and then re-run LaTeX}
  \typeout{** twice to fix the references!}
  \typeout{******************************************}
  \typeout{}
 }

\end{document}